\documentclass[preprint,preprintnumbers,amsmath,amssymb]{revtex4}

\usepackage{bm}
\usepackage{graphicx}
\usepackage{subfig}
\usepackage{array}

\graphicspath{{figures/}}

\makeatletter

\newcommand{\Rmnum}[1]{\expandafter\@slowromancap\romannumeral #1@}
\makeatother

\begin{document}

\title{Minimal modifications to the Tri-Bimaximal neutrino mixing}

\author{Zhen-hua Zhao}

\email{zhaozhenhua@ihep.ac.cn}

\affiliation{Institute of High Energy Physics, Chinese Academy of Sciences,\\
P.O. Box 918, Beijing 100049, China}

\begin{abstract}

In light of the observation of a relatively large $\theta_{13}$, the ever popular Tri-Bimaximal (TBM) neutrino mixing
which predicts a vanishing $\theta_{13}$ needs modifications.
In this paper, we shall discuss the possibility of modifying it in a minimal way to fulfil this task.
In the first part, a neutrino mass matrix with three independent parameters,
which leads to the TM2 mixing, is obtained by analogy with that for the TBM mixing.
In the second part, a model that can realize the TM2 mixing is constructed with
flavor symmetries $A_4 \times U(1) \times Z_2 \times Z_2 \times Z_2$.
It is the variant of a model that gives the TBM mixing, with only one more flavon field included.
Furthermore, the imaginary vacuum expectation value (VEV) of this flavon breaks the imposed CP symmetry
and results in $\theta_{23}=45^\circ$ and the maximal CP violation.
Besides, this model building approach can be generalized to the TM1 mixing in a straightforward way.

\end{abstract}

\maketitle

\section{Introduction}

The fact of neutrino oscillations has been established.
On the theoretical side, it can be explained by neutrinos having masses, and well described
by a $3\times3$ unitary matrix --- the PMNS matrix \cite{pmns} plus two mass squared differences
$\Delta m_{21}^2=m_2^2-m_1^2$ and $\Delta m_{31}^2=m_3^2-m_1^2$.
The PMNS matrix can be parameterized by three mixing angles and three CP phases,
\begin{equation}
U_{\rm{PMNS}}=\left(
\begin{array}{ccc}
c_{12}c_{13} & s_{12}c_{13} & s_{13}e^{-i\delta}\\
-s_{12}c_{23}-c_{12}s_{23}s_{13}e^{i\delta} & c_{12}c_{23}-s_{12}s_{23}s_{13}e^{i\delta} & s_{23}c_{13} \\
s_{12}s_{23}-c_{12}c_{23}s_{13}e^{i\delta} & -c_{12}s_{23}-s_{12}c_{23}s_{13}e^{i\delta} & c_{23}c_{13}
\end{array}
\right)
\left(
\begin{array}{ccc}
e^{i\alpha}&&\\
&e^{i\beta}&\\
&&1
\end{array}
\right),
\label{0}
\end{equation}
where $s_{ij}$ and $c_{ij}$ stand for $\sin{\theta_{ij}}$ and $\cos{\theta_{ij}}$.
As in the CKM matrix \cite{ckm}, there is a Dirac CP-violating phase $\delta$.
Differently, two Majorana phases $\alpha$ and $\beta$ may appear or not, depending on the nature of neutrino masses.
In this paper, neutrino masses will be taken as Majorana ones.
On the experimental side, neutrinos from different sources,
ranging from the sun \cite{solar} and the atmosphere \cite{atm}
to reactors \cite{dayabay} and accelerators \cite{minos},
have been observed to oscillate among different flavors.
Thanks to the accumulation of data, neutrino mixing parameters have been measured with a high precision.
According to the latest global-fit results \cite{valle}, they have the values as given in TABLE \Rmnum 1.
Only the values in the case of normal hierarchy are shown here, because the models in this paper just allow this situation.
\begin{table}[h]
\begin{tabular}{|p{3.5cm}<{\centering}|p{2.5cm}<{\centering}|p{2.5cm}<{\centering}|p{2.5cm}<{\centering}|p{2.5cm}<{\centering}|}
\hline
   Parameter      &best fit & 1 $\sigma$ range &  2  $\sigma$ range  &  3  $\sigma$ range \\
         \hline
$\sin^2{\theta_{12}}/{10^{-1}}$      &  3.23   &3.07$-$3.39 & 2.92$-$3.57  & 2.78$-$3.75 \\
\hline
$\sin^2{\theta_{13}}/{10^{-2}}$   &2.34  &2.14$-$2.54    &1.95$-$2.74      &1.77$-$2.94       \\
\hline
$\sin^2{\theta_{23}}/{10^{-1}}$  & 5.67 & 4.39$-$5.99 & 4.13$-$6.23 & 3.92$-$6.43 \\
\hline
$\Delta m_{21}^2[10^{-5}\rm{eV^2}]$ & 7.60 & 7.42$-$7.79 & 7.26$-$7.99 &7.11$-$8.18  \\
\hline
$|\Delta m_{31}^2|[10^{-3}\rm{eV^2}]$ & 2.48 & 2.41$-$2.53& 2.35$-$2.59&2.30$-$2.65\\
\hline
$\delta/\pi$  & 1.34 & 0.96$-$1.98  & 0.00$-$2.00& 0.00$-$2.00  \\
\hline
\end{tabular}
\caption{Global-fit results for neutrino oscillation parameters.}
\end{table}

Before the measurement of $\theta_{13}$ \cite{dayabay}, the TBM mixing \cite{tri-bi} was very popular,
\begin{equation}
U_{{\rm TBM}}=
\left(
\begin{array}{ccc}
\vspace{0.2cm}
\displaystyle\frac{\sqrt 2}{\sqrt 3}&\displaystyle\frac{1}{\sqrt 3}&0\\
\vspace{0.2cm}
-\displaystyle\frac{1}{\sqrt 6}&\displaystyle\frac{1}{\sqrt 3}&\displaystyle\frac{1}{\sqrt 2}\\
\vspace{0.2cm}
\displaystyle\frac{1}{\sqrt 6}&-\displaystyle\frac{1}{\sqrt 3}&\displaystyle\frac{1}{\sqrt 2}
\end{array}
\right),
\end{equation}
whose prediction for mixing angles
\begin{equation}
\sin^2{\theta_{12}}=\frac{1}{3}, \hspace{1.5cm} \sin^2{\theta_{23}}=\frac{1}{2}, \hspace{1.5cm}\theta_{13}=0,
\end{equation}
was in good agreement with experimental results at that time.
When the charged leptons are diagonal, a neutrino mass matrix of the following form can give us the TBM mixing,
\begin{equation}
M_{\nu}=
\left(
\begin{array}{ccc}
A&\hspace{0.4cm}&\hspace{0.4cm}\\
&\hspace{0.4cm}A&\hspace{0.4cm}\\
&\hspace{0.4cm}&\hspace{0.4cm}A
\end{array}
\right)
+
\left(
\begin{array}{ccc}
0&C&-C\\
C&B+C&B\\
-C&B&B+C
\end{array}
\right).
\label{111}
\end{equation}
Due to its simplicity and predictive power, many models starting from a discrete non-Abelian flavor symmetry \cite{ema} \cite{alt}  \cite{review} were proposed to realize this mass matrix and thus the TBM mixing.

However, considering the significant deviation of $\theta_{13}$ from $0$,
we need to modify the TBM mixing \cite{tm} \cite{modification}.
First of all, a natural question arises as whether there is still a neutrino mass matrix
that can accommodate the large $\theta_{13}$ and assumes a simple form like that in Eq. (\ref{111})  \cite{simple}.
In Section \Rmnum{2}, a mass matrix for this purpose is obtained through twisting Eq. (\ref{111}),
and its phenomenological consequences are discussed.
As we will see, this mass matrix actually leads to the so-called TM2 mixing \cite{tm}.
Therefore, a model with flavor symmetries is constructed to realize this mixing pattern in Section \Rmnum{3}.
Special attention will be paid to the origin of CP violation.
Furthermore, the generalization of this model to the TM1 mixing is also discussed \cite{tm} \cite{xz}.
Finally, a brief summary is given in Section \Rmnum{4}.

\section{A Minimal Modification to the Mass Matrix}

In this section, we will modify Eq. (\ref{111}) minimally to produce a realistic neutrino mixing pattern.
The mass matrix given below can take this responsibility,
\begin{equation}
M_{\nu}=\left(
\begin{array}{ccc}
a&\hspace{0.4cm}&\hspace{0.4cm}\\
&\hspace{0.4cm}a&\hspace{0.4cm}\\
&\hspace{0.4cm}&\hspace{0.4cm}a
\end{array}
\right)
+\left(
\begin{array}{ccc}
0&\hspace{0.4cm}c&\hspace{0.4cm}c\\
c&\hspace{0.4cm}b-c&\hspace{0.4cm}b\\
c&\hspace{0.4cm}b&\hspace{0.4cm}b+c
\end{array}
\right).
\label{1}
\end{equation}
It can be viewed as a sister matrix of Eq. (\ref{111}),
in the sense that their elements possess similar relations,
\begin{equation}
m_{\mu\mu}-m_{ee}=m_{\mu\tau}-m_{e\tau}, \hspace{1.5cm} m_{\tau\tau}-m_{ee}=m_{\mu\tau}+m_{e\mu}, \hspace{1.5cm} m_{e\mu}= \pm m_{e\tau}.
\label{112}
\end{equation}
The only difference lies in the fact that Eq. (\ref{111}) takes $m_{e\mu}= -m_{e\tau}$, while Eq. (\ref{1}) takes $m_{e\mu}= m_{e\tau}$.
Thus, the latter case can be taken as a minimal modification to the former case.
The above way of obtaining Eq. (\ref{1}) is a little novel and seems not to be reasonable.
However, it can also be reached from other perspectives which have solid ground.
First of all, we notice that it has something to do with the Friedberg-Lee symmetry \cite{li} \cite{fl}
which shapes the neutrino mass matrix to be as
\begin{equation}
M_{\nu}
=\left(
\begin{array}{ccc}
a&\hspace{0.4cm}&\hspace{0.4cm}\\
&\hspace{0.4cm}a&\hspace{0.4cm}\\
&\hspace{0.4cm}&\hspace{0.4cm}a
\end{array}
\right)
+
\left(
\begin{array}{ccc}
c+d&\hspace{0.4cm}-d&\hspace{0.4cm}c\\
-d&\hspace{0.4cm}b+d&\hspace{0.4cm}b\\
c&\hspace{0.4cm}b&\hspace{0.4cm}b+d
\end{array}
\right).
\end{equation}
By choosing $d=-c$, this equation can be reduced to Eq. (\ref{1}),
so the latter has one parameter fewer and its results are more predictive as we will see.
Eq. (\ref{1}) can also be understood in terms of the $\mu-\tau$ symmetry \cite{mu tau symmetry} and its breaking \cite{broken mu tau}.
As pointed out in Ref. \cite{mu tau symmetry breaking}, a general neutrino mass matrix can be decomposed into two parts,
\begin{equation}
M_{\nu}=\left(
\begin{array}{ccc}
M_{ee}&\hspace{0.3cm}M_{e\mu}^{+}&\hspace{0.3cm}-M_{e\mu}^{+}\\
M_{e\mu}^{+}&\hspace{0.3cm} M_{\mu\mu}^{+}&\hspace{0.3cm} M_{\mu\tau}\\
-M_{e\mu}^{+}&\hspace{0.3cm} M_{\mu\tau}&\hspace{0.3cm} M_{\mu\mu}^{+}
\end{array}
\right)
+
\left(
\begin{array}{ccc}
0&\hspace{0.3cm}M_{e\mu}^{-}&\hspace{0.3cm}M_{e\mu}^{-}\\
M_{e\mu}^{-}&\hspace{0.3cm} M_{\mu\mu}^{-}&\hspace{0.3cm} 0\\
M_{e\mu}^{-}&\hspace{0.3cm} 0&\hspace{0.3cm} -M_{\mu\mu}^{-}
\end{array}
\right),
\end{equation}
where the first part obeys the $\mu-\tau$ symmetry, while the second part breaks it.
In Eq. (\ref{1}), $a$ and $c$ obey the $\mu-\tau$ symmetry,
while $b$ which corresponds to taking $M_{\mu\mu}^{-}=-M_{e\mu}^{-}$ violates it.
Anyway, we can put aside the origin of Eq. (\ref{1}) for the time being and
just study its implications for phenomenology.

In order to make physical results manifest, instead of the standard parametrization in Eq. (\ref{0}),
the matrix that diagonalizes Eq. (\ref{1}) is parameterized in a different way,
\begin{equation}
U_\nu=R(\theta_{23}^\prime)R(\theta_{12}^\prime)R(\theta_{13}^\prime),
\label{2}
\end{equation}
where $R(\theta_{ij}^\prime)$ is a rotation in the i-j plane by the angle $\theta^\prime_{ij}$.
Here the superscript ``$\prime$" is used to distinguish the angles from those in the standard parametrization.
In this case, the three angles can be expressed as
\begin{equation}
\sin{\theta_{23}^\prime}=\displaystyle\frac{1}{\sqrt 2}\ , \hspace{1.5cm} \sin{\theta_{12}^\prime}=\displaystyle\frac{1}{\sqrt 3}\ , \hspace{1.5cm}
\sin{\theta_{13}^\prime}=\displaystyle\frac{\sqrt{b^2+3c^2}-b^2}{\sqrt{2(b^2+3c^2)-2b^2\sqrt{b^2+3c^2}}}\ ,
\label{4}
\end{equation}
while the mass eigenvalues are
\begin{equation}
m_1=a+b-\sqrt{b^2+3c^2}\ , \hspace{1.5cm} m_2=a\ , \hspace{1.5cm} m_3=a+b+\sqrt{b^2+3c^2}\ .
\label{5}
\end{equation}
In the basis where the charged leptons are diagonal, $U_{\rm{PMNS}}$ is identical with $U_{\nu}$,
\begin{equation}
U_{\rm{PMNS}}=\left(
\begin{array}{ccc}
\vspace{0.2cm}
\displaystyle\frac{\sqrt 2}{\sqrt 3}&\displaystyle\frac{1}{\sqrt 3}&0\\
\vspace{0.2cm}
-\displaystyle\frac{1}{\sqrt 6}&\displaystyle\frac{1}{\sqrt 3}&\displaystyle\frac{1}{\sqrt 2}\\
\vspace{0.2cm}
\displaystyle\frac{1}{\sqrt 6}&-\displaystyle\frac{1}{\sqrt 3}&\displaystyle\frac{1}{\sqrt 2}
\end{array}
\right)
\left(
\begin{array}{ccc}
\vspace{0.2cm}
\cos{\theta_{13}^\prime} &   &  \sin{\theta_{13}^\prime}  \\
\vspace{0.2cm}
& 1  &  \\
\vspace{0.2cm}
 -\sin{\theta_{13}^\prime}  &  & \cos{\theta_{13}^\prime}
\end{array}
\right).
\label{6}
\end{equation}
Confronting Eq. (\ref{6}) with Eq. (\ref{0}), neutrino mixing angles in the standard parameterization
can be extracted as follows,
\begin{equation}
\sin{\theta_{13}}= \displaystyle{\frac{\sqrt2}{\sqrt3}} \sin{\theta_{13}^\prime}\ ,  \hspace{0.6cm}
\sin{\theta_{12}}= \frac{{\displaystyle\frac{1}{\sqrt3}}}{\sqrt{1-\displaystyle\frac{2}{3} \sin^2{\theta_{13}^\prime} }}\ ,\hspace{0.6cm}
\sin{\theta_{23}}=\displaystyle\frac{(-{\displaystyle\frac{1}{\sqrt6}} \sin{\theta_{13}^\prime} +{\displaystyle\frac{1}{\sqrt2}} \cos{\theta_{13}^\prime} )}{\sqrt{1-\displaystyle\frac{2}{3} \sin^2{\theta_{13}^\prime}}}\ .
\label{9}
\end{equation}
For definiteness, $\theta_{13}^\prime$ will take the following value which gives $\sin{\theta_{13} }=0.15$,
\begin{equation}
\sin{\theta_{13}^\prime}=0.19\ , \hspace{1.5cm}  {\rm{when}} \hspace{1cm} \displaystyle\frac{c}{b}=\frac{2}{5\sqrt 3}\ .
\label{13prime}
\end{equation}
With this choice, the mixing angles can be calculated directly,
\begin{equation}
\sin^2{\theta_{13}}= 0.0237, \hspace{1.5cm}
\sin^2{\theta_{12}}= 0.341, \hspace{1.5cm}
\sin^2{\theta_{23}}= 0.390\ .
\label{mixing angles}
\end{equation}
Furthermore, the values of $a$, $b$ and $c$ are completely determined,
\begin{equation}
a=5.06\times 10^{-2}{\rm{eV}}, \hspace{1.5cm} b=9.81\times 10^{-3}{\rm{eV}}, \hspace{1.5cm} c=2.27\times10^{-3}{\rm{eV}}.
\end{equation}
As a result, neutrino masses are calculable and they are of the normal hierarchy
\begin{equation}
m_1=4.98 \times 10^{-2}{\rm{eV}}, \hspace{1.5cm} m_2=5.06\times 10^{-2}{\rm{eV}}, \hspace{1.5cm} m_3=7.10\times10^{-2}{\rm{eV}}.
\end{equation}
$m_{\beta\beta}$ which regulates the rate of neutrino-less double beta decay
and $\sum m_i$ can be obtained as
\begin{equation}
m_{\beta\beta}=0.051{\rm{eV}}, \hspace{1.5cm}  \sum m_i=0.171{\rm{eV}},
\label{masses}
\end{equation}
which are very close to the experimental upper bounds, so expected to be observable in the near future.

The result for $\theta_{23}$ is on the edge of the 3 $\sigma$ range of the global-fit results
and outside of the 2 $\sigma$ range of T2K's recent result $\sin^2{\theta_{23}}=0.514^{+0.055}_{-0.056}$ \cite{t2k}.
However, it is consistent with results of the MINOS experiment $\sin^2{\theta_{23}}=0.388^{+0.051}_{-0.035} $ \cite{minos}
or $\sin^2{\theta_{23}}=0.35-0.65$ ($90\%$ C.L.) in another analysis \cite{minos2}.
Thus, we can't come to a definite conclusion before $\theta_{23}$ is measured with a high precision.
More importantly, the prediction for $\theta_{23}$ will be changed if CP violation is taken into consideration.
For example, we can take $c$ as a complex parameter $|c|e^{i\phi}$ while keeping $a$ and $b$ real,
for which case the PMNS matrix becomes,
\begin{equation}
U_{{\rm PMNS}}=
\left(
\begin{array}{ccc}
\vspace{0.2cm}
\displaystyle\frac{\sqrt 2}{\sqrt 3}&\displaystyle\frac{1}{\sqrt 3}&0\\
\vspace{0.2cm}
-\displaystyle\frac{1}{\sqrt 6}&\displaystyle\frac{1}{\sqrt 3}&\displaystyle\frac{1}{\sqrt 2}\\
\vspace{0.2cm}
\displaystyle\frac{1}{\sqrt 6}&-\displaystyle\frac{1}{\sqrt 3}&\displaystyle\frac{1}{\sqrt 2}
\end{array}
\right)
\left(
\begin{array}{ccc}
\vspace{0.2cm}
\cos{\theta_{13}^\prime}&  &\sin{\theta_{13}^\prime}e^{-i \rho}\\
\vspace{0.2cm}
 &1&\\
 \vspace{0.2cm}
-\sin{\theta_{13}^\prime}e^{i \rho}& &\cos{\theta_{13}^\prime}
\end{array}
\right),
\label{tm2}
\end{equation}
where $\theta_{13}^\prime$ and $\rho$ can be obtained from
\begin{equation}
\tan{\rho}=\displaystyle\frac{b\tan{\phi} }{a+b},  \hspace{1.5cm} \tan{2 \theta_{13}^\prime}=\displaystyle\frac{\sqrt{3}c \cos{\phi}}{b \cos{\rho}}.
\end{equation}
Accordingly, there is a correlation among $\theta_{13}$, $\theta_{23}$ and $\rho$
\begin{equation}
\sin^2{\theta_{23}}\hspace{0.15cm}=\hspace{0.15cm}\displaystyle\frac{\displaystyle\frac{1}{2}(1-\sin^2{\theta_{13}})-\displaystyle\frac{1}{\sqrt 3}\cos{\rho}\sin{\theta_{13}}\sqrt{\displaystyle\frac{3}{2}(1-\displaystyle\frac{3}{2}\sin^2{\theta_{13}})}}{1-\sin^2{\theta_{13}}}.
\label{theta23}
\end{equation}
For illustration, we can fix $\sin{\theta_{13}}$ at 0.15,
then $\sin^2{\theta_{23}}$ would vary from 0.393 to 0.607 when $\rho$ takes values in the range $[0, 2 \pi]$.
Obviously, we can go back to the mixing matrix given in Eq. (\ref{6}) by taking $\rho=0$.
On the other hand, in the case of $\rho=\pi/2$ or $3\pi/2$, $\theta_{23}$ remains maximal.
This interesting possibility \cite{tri} is still allowed by experimental results
and provides a promising CP-violating effect, with the Jarlskog invariant \cite{jarlskog} as large as 0.036.

\section{A Minimal Modification to the Model Building}

As we have seen, the modified mass matrix Eq. (\ref{1}) results in the TM2 mixing given by Eq. (\ref{tm2})
where the particular case $\rho=\pi/2 {\rm or} 3\pi/2$ deserves special attention.
A model realizing this mixing pattern will be given in the following.
There have already been several models for this purpose in the literature \cite{tm2} \cite{review2}.
But from a different point of view, we will achieve this goal by modifying a model that gives the TBM mixing as minimally as possible.
Our starting point is an observation:
As Eq. (\ref{tm2}) itself suggests, the PMNS matrix can be split into two parts which have different origins.
This can be realized through the following thread:
At the first stage, right handed neutrinos $N_1, N_2$ and $N_3$ are diagonal and their Yukawa couplings with
left-handed neutrinos have such a form that light neutrinos have the TBM mixing after the seesaw mechanism \cite{seesaw}.
At the second stage, a flavon field which acquires a VEV induces the mixing between $N_1$ and $N_3$,
contributing the second part of the mixing matrix.
In order to control the source of CP violation, we will impose the CP symmetry \cite{spontaneous}
and spontaneously break it by this same flavon field.

The model is constructed under the simplest non-Abelian discrete group --- $A4$,
which has 4 different representations ${\bf1}, {\bf1^{\prime}}, {\bf1^{\prime \prime}}$ and ${\bf3}$,
whose multiplication rules are listed here for consultation \cite{a4},
\begin{equation}
\begin{array}{lllllll}
1^\prime &\times & 1^{\prime \prime}&\hspace{0.3cm} \rightarrow & \hspace{0.3cm}1&=&a b\\
1^\prime &\times & 3 &\hspace{0.3cm}\rightarrow & \hspace{0.3cm} 3 &=&(a b_3, a b_1, a b_2)\\
1^{\prime \prime}  &\times & 3 &\hspace{0.3cm}\rightarrow & \hspace{0.3cm} 3&=&(a b_2, a b_3, a b_1)\\
3 &\times & 3  &\hspace{0.3cm}\rightarrow & \hspace{0.3cm} 1&=& a_1 b_1+ a_2 b_3 + a_3 b_2\\
3&\times & 3  &\hspace{0.3cm}\rightarrow & \hspace{0.3cm} 1^\prime &=& a_3 b_3+ a_1 b_2 + a_2 b_1\\
3&\times &3  &\hspace{0.3cm}\rightarrow & \hspace{0.3cm} 1^{\prime \prime} &=& a_2 b_2+ a_1 b_3 + a_3 b_1\\
3&\times & 3  &\hspace{0.3cm}\rightarrow & \hspace{0.3cm} 3_A&=& ( a_2 b_3 - a_3 b_2,  a_1 b_2 - a_2 b_1, a_1 b_3 - a_3 b_1)\\
\end{array}
\end{equation}
and
\begin{equation}
3\times 3 \rightarrow  3_S= (2 a_1 b_1 - a_2 b_3 - a_3 b_2, 2 a_3 b_3 - a_1 b_2 - a_2 b_1, 2 a_2 b_2 - a_1 b_3 - a_3 b_1).
\end{equation}
$a_i$ and $b_i$ denote the components of a multi-dimensional representation.
In order to establish the relations among different mass matrix elements as suggested by Eq. (\ref{112}),
three lepton doublets $L_{i=1,2,3}$ are organized to form the representation ${\bf3}$.
Since there are large hierarchies among the charged leptons, $e^c$, $\mu^c$ and $\tau^c$
(here we have employed the convention in supersymmetry(SUSY) to denote the singlets under the $SU(2)_{{\rm L}}$ gauge symmetry)
are specified as representations ${\bf1}$, ${\bf1^{\prime \prime}}$ and ${\bf1^\prime}$ respectively.
An additional $U(1)$ symmetry, which plays the same role as the well-known Froggatt-Nielsen symmetry \cite{fn},
is introduced to produce these hierarchies, by letting $e^c$, $\mu^c$ and $\tau^c$ have different charges under it.
As mentioned, we want the mass matrix for right-handed neutrinos to be diagonal at the first step,
so they are arranged to be the representation ${\bf1}$ and have $Z_2^{i=1,2,3}$ quantum numbers respectively.
The flavon field $\xi$ which is charged under both $Z_2^1$ and $Z_2^3$ will induce the mixing between $N_1$ and $N_3$ after obtaining a VEV.
Finally, there are some other flavon fields $\phi$, $\varphi$, $\chi$ and $\psi$ which will spontaneously break the $A4$ symmetry.
All the fields and their quantum numbers are summarized in TABLE \Rmnum 2.

\begin{table}[h]
\begin{tabular}{|p{0.8cm}<{\centering}|p{0.8cm}<{\centering}|p{0.8cm}<{\centering}|p{0.8cm}<{\centering}|p{0.8cm}<{\centering}|p{0.8cm}<{\centering}|p{0.8cm}<{\centering}|p{0.8cm}<{\centering}|p{0.8cm}<{\centering}|p{0.8cm}<{\centering}|p{0.8cm}<{\centering}|p{0.8cm}<{\centering}|p{0.8cm}<{\centering}|p{0.8cm}<{\centering}|}
\hline
         &$L_i$ &$e^c$ &$\mu^c$&$\tau^c$ &$N_1$ &$N_2$ & $N_3$ &$H_{{{\rm u,d}}}$ &$\phi$    &$\varphi$   &$\chi$ &$\psi $  & $ \xi $     \\
         \hline
$A4$     &3      &1  &$1^{\prime\prime}$ &$1^{\prime}$  &1 &1 &1    &1     &3        &3         &3    &3       &1 \\
\hline
$U1$    &1  &-5    &-3                &-2    &0          &0     &0 &0     &1         &-1      &-1   & -1&   0     \\
\hline
$Z_{2}^1$  & 1 & 1& 1&1 &-1 &1  &1 & 1 & 1& -1& 1&1 &-1  \\
\hline
$Z_{2}^2$  & 1 & 1& 1&1 &1 &-1  &1 & 1& 1& 1& -1&1 &1  \\
\hline
$Z_{2}^3$ & 1 & 1& 1&1 &1 &1  &-1 & 1& 1& 1& 1&-1 &-1  \\
\hline
\end{tabular}
\caption{Quantum numbers of the fields.}
\end{table}

\subsection{The VEV Alignments}

In models with discrete flavor symmetries, flavon fields such as $\phi$, $\varphi$, $\chi$ and $\psi$
which are multi-dimensional representations are normally required to have VEVs with specific alignments,
so that a particular mixing pattern can be guaranteed.
This model is not an exception and the VEVs have a form as follows
\begin{equation}
\langle \phi \rangle=(1,0,0)V_{1}, \hspace{0.7cm} \langle \varphi \rangle=(2,-1,-1)V_2,\hspace{0.7cm}
\langle \chi \rangle=(1,1,1)V_{3}, \hspace{0.7cm} \langle \psi \rangle =(0,1,-1)V_4.
\label{vev}
\end{equation}
As usual, the reasonableness of this choice can be justified by the approach developed in \cite{alt}:
In the framework of SUSY, we can introduce some ``driving fields" to make the flavon fields
have the required VEVs, with the help of R-symmetry --- $U(1)_{{\rm R}}$.
Driving fields are the ones that have charge 2 under $U(1)_{{\rm R}}$, while the flavon fields have charge 0.
Since the terms in superpotential are required to have charge 2 in total, they should have the form $\Delta(\cdots)$,
where $\Delta$ represents a driving field and dots in the bracket are linear combinations of the flavon fields.
If there are driving fields with quantum numbers as shown in TABLE \Rmnum 3,
the superpotential --- $\mathcal W$, which is relevant to VEVs of the flavon fields,
is constrained to the following form, up to next-to-leading-order (NLO),
\begin{equation}
\begin{array}{l}
\vspace{0.2cm}
\Delta_1\{\lambda_1(\phi \phi)_{1^{\prime \prime}}\}+\Delta_2\{\lambda_2(\phi \phi)_{1^{\prime}}\}
+\Delta_3\{\displaystyle\frac{\lambda_3}{\Lambda}(\phi \phi \psi )_{1}\}+ \Delta_4\{\lambda_4(\xi \xi)_{1} \pm M^2\}+\\
\vspace{0.2cm}
\Delta_5\{\lambda_5(\varphi \varphi)_{1}+\lambda_6(\chi \chi)_{1}+\lambda_{7}(\psi \psi)_{1}+\displaystyle\frac{\lambda_{8}}{\Lambda}(\varphi \psi \xi)_1\}+\Delta_6\{ \lambda_{9} (\varphi \chi)_1
+\displaystyle\frac{\lambda_{10}}{\Lambda}(\chi \psi \xi)_1 \}+\\
\vspace{0.2cm}
\Delta_7\{ \lambda_{11} (\varphi \chi)_{1^{\prime \prime}}+\displaystyle\frac{\lambda_{12}}{\Lambda}(\chi \psi \xi)_{1^{\prime \prime}} \}
+\Delta_8\{ \lambda_{13} (\varphi \chi)_{1^{\prime}}+\displaystyle\frac{\lambda_{14}}{\Lambda}(\chi \psi \xi)_{1^{\prime}} \}+\Delta_{9}\{ \lambda_{15} (\varphi \psi)_{1}+\\
\vspace{0.2cm}
\displaystyle\frac{\lambda_{16}}{\Lambda}(\varphi \varphi \xi)_{1}+\displaystyle\frac{\lambda_{17}}{\Lambda}(\chi \chi \xi)_{1}+\displaystyle\frac{\lambda_{18}}{\Lambda}(\psi \psi \xi)_{1}\}
+\Delta_{10}\{\lambda_{19} (\chi \psi)_1+\displaystyle\frac{\lambda_{20}}{\Lambda}(\varphi \chi \xi)_1 \}+\\
\vspace{0.2cm}
\Delta_{11}\{\lambda_{21} (\chi \psi)_{1^{\prime \prime}}+\displaystyle\frac{\lambda_{22}}{\Lambda}(\varphi \chi \xi)_{1^{\prime \prime}} \} +\Delta_{12}+\{\lambda_{23} (\chi \psi)_{1^{\prime}}+\displaystyle\frac{\lambda_{24}}{\Lambda}(\varphi \chi \xi)_{1^{\prime}} \}.
\end{array}
\end{equation}
In the above, $\lambda_i$ are dimensionless coefficients and $M$ is a dimension-one parameter,
and $\Lambda$ is the cut-off scale for non-renormalizable operators.
The symbol $(\cdots)_{1/1^\prime/1^{\prime \prime}}$ means that linear combinations in the bracket must form
the representation ${\bf1}$ or ${\bf1^\prime}$ or ${\bf1^{\prime \prime}}$ to match the corresponding $\Delta_i$.

\begin{table}[h]
\begin{tabular}{|p{0.8cm}<{\centering}|p{0.8cm}<{\centering}|p{0.8cm}<{\centering}|p{0.8cm}<{\centering}|p{0.8cm}<{\centering}|p{0.8cm}<{\centering}|p{0.8cm}<{\centering}|p{0.8cm}<{\centering}|p{0.8cm}<{\centering}|p{0.8cm}<{\centering}|p{0.8cm}<{\centering}|p{0.8cm}<{\centering}|p{0.8cm}<{\centering}|}
\hline
         &$\Delta_1$ &$\Delta_2$ &$\Delta_3$&$\Delta_4$  &$\Delta_5$ & $\Delta_6$ &$\Delta_7$ &$\Delta_8$    &$\Delta_{9}$   &$\Delta_{10}$ &$\Delta_{11} $  & $\Delta_{12} $     \\
         \hline
$A4$  &$1^{\prime}$ &$1^{\prime\prime}$ &1 &1  &1 &1 &$1^{\prime}$   &$1^{\prime\prime}$  &1   &1    &$1^{\prime}$   &$1^{\prime\prime}$ \\
\hline
$U1$    &-2  &-2    &-1               &0            &2     &2 &2     &2         &2      &2   & 2&  2     \\
\hline
$Z_{2}^1$  & 1 & 1& 1&1  &1  &-1 & -1 & -1& -1& 1&1 &1  \\
\hline
$Z_{2}^2$  & 1 & 1& 1&1 &1  &-1 & -1& -1& 1& -1&-1 &-1  \\
\hline
$Z_{2}^3$  & 1 & 1& -1&1  &1  &1 & 1& 1& -1& -1&-1 &-1  \\
\hline
\end{tabular}
\caption{Quantum numbers of Driving Fields.}
\end{table}

SUSY requires each $F$ component of the driving fields to have a vanishing VEV,
\begin{equation}
\langle F_i^* \rangle= - \frac{\partial {\mathcal W}}{\partial \Delta_i}=0.
\end{equation}
This leads to some constraint equations on the VEVs,
\begin{equation}
\begin{array}{llll}
\lambda_1(\phi_2 \phi_2 + 2 \phi_1 \phi_3)&=  0,&\hspace{1cm}
\lambda_2(\phi_3 \phi_3 + 2 \phi_1 \phi_2)& = 0,\\
\lambda_{8}(\varphi_1 \psi_1 + \varphi_2 \psi_3+\varphi_3 \psi_2)\xi& =0,&\hspace{1cm}
\lambda_{9}(\varphi_1 \chi_1 +\varphi_2 \chi_3 +\varphi_3 \chi_2)&=0,\\
\lambda_{10}(\chi_1 \psi_1+ \chi_2 \psi_3+ \chi_3 \psi_2)\xi&=0,&\hspace{1cm}
\lambda_{11}(\varphi_2 \chi_2 +\varphi_1 \chi_3 +\varphi_3 \chi_1)&=0,\\
\lambda_{12}(\chi_2 \psi_2+ \chi_1 \psi_3+ \chi_3 \psi_1)\xi&=0,&\hspace{1cm}
\lambda_{13}(\varphi_3 \chi_3 +\varphi_1 \chi_2 +\varphi_2 \chi_1)&=0,\\
\lambda_{14}(\chi_3 \psi_3+ \chi_1 \psi_2+ \chi_2 \psi_1)\xi&=0,&\hspace{1cm}
\lambda_{15}(\varphi_1 \psi_1 +\varphi_2 \psi_3+\varphi_3 \psi_2)&=0,\\
\lambda_{19}(\chi_1 \psi_1 +\chi_2 \psi_3 +\chi_3 \psi_2)&=0,&\hspace{1cm}
\lambda_{20}(\varphi_1 \chi_1+ \varphi_2 \chi_3+ \varphi_3 \chi_2)\xi&=0,\\
\lambda_{21}(\chi_2 \psi_2 +\chi_1 \psi_3 +\chi_3 \psi_1)&=0,&\hspace{1cm}
\lambda_{22}(\varphi_2 \chi_2+ \varphi_1 \chi_3+ \varphi_3 \chi_1)\xi&=0,\\
\lambda_{23}(\chi_3 \psi_3 +\chi_1 \psi_2 +\chi_2 \psi_1)&=0,&\hspace{1cm}
\lambda_{24}(\varphi_3 \chi_3+ \varphi_1 \chi_2+ \varphi_2 \chi_1)\xi&=0,\\
\end{array}
\end{equation}
and
\begin{equation}
\begin{array}{ll}
\lambda_3[(\phi_1 \phi_1 -\phi_2 \phi_3)\psi_1+ (\phi_3 \phi_3 -\phi_1 \phi_2)\psi_3+(\phi_2 \phi_2 -\phi_1 \phi_3)\psi_2]& =0,\\
\lambda_5(\varphi_1 \varphi_1+ 2\varphi_2 \varphi_3)+ \lambda_6(\chi_1 \chi_1+ 2\chi_2 \chi_3)+ \lambda_{7}(\psi_1 \psi_1+ 2\psi_2 \psi_3)&=0,\\
\lambda_{16}(\varphi_1 \varphi_1+ 2\varphi_2 \varphi_3) \xi+\lambda_{17}(\chi_1 \chi_1+ 2\chi_2 \chi_3) \xi+\lambda_{18}(\psi_1 \psi_1+ 2\psi_2 \psi_3) \xi&=0.\\
\end{array}
\label{ab}
\end{equation}
Eq. (\ref{vev}) is a solution to these equations,
so it is fair to say that the VEVs can have the form as shown by it at least to NLO.
Besides, there are some relations among $V_2,V_3$ and $V_4$,
\begin{equation}
V_3=\sqrt{\frac{2\lambda_5 \lambda_{18}-2 \lambda_7 \lambda_{16}}{\lambda_7 \lambda_{17}-\lambda_6 \lambda_{18}}}V_2, \hspace{1cm}
V_4=\sqrt{\frac{3\lambda_5 \lambda_{17}-3 \lambda_6 \lambda_{16}}{\lambda_7 \lambda_{17}-\lambda_6 \lambda_{18}}}V_2.
\label{cd}
\end{equation}
In particular, $\xi$ also gets a VEV: $V_5=\pm M/\sqrt{\lambda_4}$ for the minus sign
or $V_5=\pm i M/\sqrt{\lambda_4}$ for the plus sign in the below equation,
\begin{equation}
\lambda_4 \xi \xi \pm M^2 =0.
\end{equation}
The latter case will be the only source for CP violation in the lepton sector, if the CP symmetry is required.
This method of obtaining a complex VEV for a scalar field is proposed in \cite{vev}.

\subsection{The Mass Matrix and Mixing Pattern}

Now we can discuss the consequences of the model on mass matrices and the mixing pattern.
The flavor symmetries only allow higher than dimension-four Yukawa-like operators,
which have a general form $y_{ij} N_i^c L_j H_{{\rm u}} (\Phi/\Lambda)^{n_{ij}}$ or $y_{ij} E_i^c L_j H_{{\rm d}} (\Phi/\Lambda)^{n_{ij}}$.
$\Phi$ represents the flavon fields $\phi, \varphi, \chi, \psi$ and $\omega$,
and $\Lambda$ is the cut-off scale where an underlying theory emerges,
when $n_{ij}$ is an integer measuring the power of $\Phi/\Lambda$.
After the flavon fields gain VEVs which are commonly denoted as $V$, these operators become effective Yukawa terms
$y_{ij} N_i^c L_j H_{{\rm u}} (V/\Lambda)^{n_{ij}}$ or $y_{ij}E^c_i L_j H_{{\rm d}}  (V/\Lambda)^{n_{ij}}$.
Usually, $V/\Lambda$ (labeled as $\epsilon$) is assumed to be an $\mathcal O(0.1)$ quantity.
In this case, effective Yukawa couplings $y_{ij}(V/\Lambda)^{n_{ij}}$ are controlled by the corresponding coefficients --- $\epsilon^{n_{ij}}$,
so that mass hierarchies can be understood in terms of the power of $\epsilon$.

The terms that contribute to masses of the charged leptons include
\begin{equation}
y_1 \tau^c L H_{{\rm d}}  \frac{\phi}{\Lambda}+y_2 \mu^c L H_{{\rm d}}  \frac{\phi^2}{\Lambda^2}+y_3 e^c L H_{{\rm d}} \frac{\phi^4}{\Lambda^4},
\end{equation}
which lead to a diagonal mass matrix
\begin{equation}
M_{{\rm l}}=\left(
\begin{array}{ccc}
y_3 \displaystyle\left(\frac{V_1}{\Lambda}\right)^4&&\\
&y_2 \displaystyle\left(\frac{V_1}{\Lambda}\right)^2&\\
&&y_1 \displaystyle\frac{V_1}{\Lambda}
\end{array}
\right)V_{{\rm d}}
\hspace{0.3cm}
\sim
\hspace{0.3cm}
\left(
\begin{array}{ccc}
\epsilon_1^4&\hspace{0.3cm}&\hspace{0.3cm}\\
&\hspace{0.3cm}\epsilon_1^2&\hspace{0.3cm}\\
&\hspace{0.3cm}&\hspace{0.3cm}\epsilon_1
\end{array}
\right)V_{{\rm d}},
\end{equation}
where $V_{{\rm d}}$ is the VEV of $H_{{\rm d}}$ and $V_1/\Lambda$ is replaced with $\epsilon_1$, indicating that it is a small quantity.
Accordingly, the hierarchies of $m_e$, $m_\mu$ and $m_\tau$ get an explanation.
On the other side, $Z_2^{i=1,2,3}$ symmetries fix the mass matrix for $N_1,N_2$ and $N_3$ to be diagonal too,
\begin{equation}
M_{{\rm N}}=\left(
\begin{array}{ccc}
M_1&&\\
&M_2&\\
&&M_3
\end{array}
\right),
\end{equation}
which arises from the Majorana mass terms
\begin{equation}
 M_1 N_1^c N_1^c + M_2 N_2^c N_2^c + M_3 N_3^c N_3^c.
\end{equation}
The flavon fields $\varphi,\chi,\psi$ which have quantum numbers separately under $Z_2^{i=1,2,3}$
make the Yukawa couplings between left-handed and right-handed neutrinos have a form
\begin{equation}
y_4 N_1^c L H_{{\rm u}}  \frac{\varphi}{\Lambda}+ y_5 N_2^c L H_{{\rm u}}  \frac{\chi}{\Lambda} + y_6  N_3^c L H_{{\rm u}}\frac{\psi}{\Lambda},
\end{equation}
which give the Dirac neutrino mass matrix as
\begin{equation}
M_{{\rm D}}=\left(
\begin{array}{ccc}
2 y_4&\hspace{0.35cm} -y_4&\hspace{0.35cm} -y_4\\
y_5^\prime&\hspace{0.35cm} y_5^\prime &\hspace{0.35cm}  y_5^\prime\\
0&\hspace{0.35cm} -y_6^\prime&\hspace{0.35cm}  y_6^\prime
\end{array}
\right) \epsilon_2 V_{{\rm u}},
\label{md}
\end{equation}
where $V_{{\rm u}}$ is the VEV of $H_{{\rm u}}$ and $\epsilon_2$ is $V_2/\Lambda$,
$y_5^\prime$ and $y_6^\prime$ are the abbreviations for $y_5\sqrt{(2\lambda_5 \lambda_{18}-2 \lambda_7 \lambda_{16})/(\lambda_7 \lambda_{17}-\lambda_6 \lambda_{18})}$ and $y_6\sqrt{(3\lambda_5 \lambda_{17}-3 \lambda_6 \lambda_{16})/(\lambda_7 \lambda_{17}-\lambda_6 \lambda_{18})}$.

The mass matrix for light neutrinos can be obtained through the seesaw mechanism,
\begin{equation}
\begin{array}{ll}
\vspace{0.25cm}
M_\nu &=M_{{\rm D}}^{{\rm T}} M_{{\rm N}}^{-1}M_{{\rm D}}\\
&=\left(
\begin{array}{ccc}
\vspace{0.25cm}
\displaystyle\frac{4 y_4^2}{M_1}+\displaystyle\frac{y_5^{\prime 2}}{M_2}&\hspace{0.35cm}\displaystyle\frac{y_5^{\prime 2}}{M_2}-\displaystyle \frac{2 y_4^2}{M_1} &\hspace{0.35cm}\displaystyle\frac{y_5^{\prime 2}}{M_2}-\displaystyle\frac{2 y_4^2}{M_1}\\
\vspace{0.25cm}
\cdots&\hspace{0.35cm}\displaystyle \frac{ y_4^2}{M_1}+\frac{y_5^{\prime 2}}{M_2} +\frac{y_6^{\prime 2}}{M_3}&\hspace{0.35cm}\displaystyle\frac{ y_4^2}{M_1}+\frac{y_5^{\prime 2}}{M_2}-\frac{y_6^{\prime 2}}{M_3}\\
\vspace{0.25cm}
\cdots&\hspace{0.35cm}\cdots &\hspace{0.35cm}\displaystyle \frac{ y_4^2}{M_1}+\frac{y_5^{\prime 2}}{M_2} +\frac{y_6^{\prime 2}}{M_3}
\end{array}
\right) (\epsilon_2 V_{{\rm u}} )^2,
\end{array}
\end{equation}
where the elements represented by $``\cdots"$ can be known through the symmetric property of $M_\nu$.
It can be diagonalized by the TBM matrix
\begin{equation}
U_{{\rm TBM}}^{{\rm T}} M_\nu U_{{\rm TBM}}=\left(
\begin{array}{ccc}
\displaystyle\frac{6 y_4^2}{M_1}&\hspace{0.35cm} &\hspace{0.35cm}\\
&\hspace{0.35cm}\displaystyle\frac{3 y_5^{\prime 2}}{M_2}&\hspace{0.35cm}\\
&\hspace{0.35cm} & \hspace{0.35cm}\displaystyle\frac{2 y_6^{\prime 2}}{M_3}
\end{array}
\right)(\epsilon_2 V_{{\rm u}} )^2.
\end{equation}
Both of the normal and inverted hierarchies are allowed by this mass spectrum, but it's more natural for the former case.
This is because the latter case needs a fine-tuning at one percent level to make the first two mass eigenvalues nearly degenerate,
but $6y_4^2/M_1$ and $3 y_5^{\prime 2}/M_2$ are two independent quantities.
With the assumption $6 y_4^2 \sim 3 y_5^{\prime 2} \sim 2 y_6^{\prime 2}$, the hierarchy between $\Delta m_{21}^2$ and $\Delta m_{31}^2$
can be attributed to the hierarchies among right-handed neutrinos $M_1 \sim M_2 \sim 5 M_3$.
Up to now, we have reproduced the well-known TBM mixing.
In the following, we will obtain the TM2 mixing by including the effect of $\xi$.

After $\xi$ obtains a VEV, the mass matrix for $N_1, N_2$ and $N_3$ becomes
\begin{equation}
M_{{\rm N}}^\prime=\left(
\begin{array}{ccc}
M_1&& \Delta M\\
&M_2&\\
\Delta M& &M_3
\end{array}
\right),
\label{mn}
\end{equation}
where $\Delta M=y V_5$ comes from the term $y N_1 N_3 \hspace{0.1cm}\xi$.
$M_{\rm D}$ remains the form given by Eq. (\ref{md}), so the mass matrix for light neutrinos turns into
\begin{equation}
\begin{array}{l}
\vspace{0.3cm}
M_{\nu}^\prime= M_{{\rm D}}^{{\rm T}} M_{{\rm N}}^{\prime -1} M_{{\rm D}}=(\epsilon_2 V_{{\rm u}} )^2\\
{\bf \cdot}\left(
\begin{array}{ccc}
\vspace{0.2cm}
\displaystyle\frac{y_5^{\prime 2}}{M_2}+\displaystyle\frac{4 y_4^2 M_3}{D}
& \displaystyle\frac{y_5^{\prime 2}}{M_2}+\displaystyle\frac{2 y_4 (y_6^\prime \Delta M- y_4 M_3)}{D}
& \displaystyle\frac{y_5^{\prime 2}}{M_2}- \displaystyle\frac{2 y_4 (y_6^\prime \Delta M+ y_4 M_3)}{D}\\
\vspace{0.2cm}
\cdots
&\displaystyle\frac{y_5^{\prime 2}}{M_2}+\displaystyle\frac{y_6^{\prime 2}M_1-2y_4 y_6^\prime \Delta M+ y_4^2 M_3}{D}
&\displaystyle\frac{y_5^{\prime 2}}{M_2} -\displaystyle\frac{y_6^{\prime 2}M_1-y_4^2 M_3}{D}\\
\vspace{0.2cm}
\cdots
&\cdots
& \displaystyle\frac{y_5^{\prime 2}}{M_2}+\displaystyle\frac{y_6^{\prime 2}M_1+2y_4 y_6^\prime \Delta M+ y_4^2 M_3}{D}
\end{array}
\right),
\end{array}
\end{equation}
where $D=M_1 M_3-(\Delta M)^2$.
A TBM rotation transforms it into the following form
\begin{equation}
U_{{\rm TBM}}^{{\rm T}} M_\nu^\prime U_{{\rm TBM}}=\left(
\begin{array}{ccc}
\displaystyle\frac{6 y_4^2 M_3}{D}&  &\displaystyle -\frac{2 \sqrt{3} y_4 y_6^\prime \Delta M}{D}\\
&\displaystyle \frac{3y_5^{\prime 2}}{M_2} &\\
 \cdots&
& \displaystyle \frac{ 2 y_6^{\prime 2} M_1}{D}
\end{array}
\right)(\epsilon_2 V_{{\rm u}} )^2,
\label{theta13}
\end{equation}
which is then diagonalized by a rotation in the 1-3 plane,
\begin{equation}
U(\theta_{13}^\prime)=\left(
\begin{array}{ccc}
\cos{\theta_{13}^\prime}&& \sin{\theta_{13}^\prime}e^{-i\rho}\\
&1&\\
-\sin{\theta_{13}^\prime}e^{i\rho}& &\cos{\theta_{13}^\prime}
\end{array}
\right),
\end{equation}
with
\begin{equation}
\begin{array}{llll}
\vspace{0.3cm}
\rho=0, & \hspace{0.6cm} \tan{2\theta_{13}^\prime}= \displaystyle\frac{2\sqrt 3 y_4 y_6^\prime \Delta M}{3 y_4^2 M_3-y_6^{\prime 2}M_1}, &\hspace{1.2cm} \rm{if} &\hspace{0.6cm} V_5=\pm \displaystyle\frac{M}{\sqrt{\lambda_4}}; \\
\rho=\displaystyle \frac{3\pi}{2}/\frac{\pi}{2},  & \hspace{0.6cm} \tan{2\theta_{13}^\prime}=\displaystyle\frac{2\sqrt 3 y_4 y_6^\prime |\Delta M|}{3 y_4^2 M_3+y_6^{\prime 2}M_1},&\hspace{1.2cm} \rm{if} &\hspace{0.6cm} V_5=\pm i\displaystyle\frac{M}{\sqrt{\lambda_4}}.
\end{array}
\end{equation}
As a result, the PMNS matrix $U_{{\rm PMNS}}=U_{{\rm TBM}}U(\theta_{13}^\prime)$ has the same form as the TM2 mixing
in three special cases: $\rho=0$, $\rho=\pi/2$ or $\rho=3\pi/2$.
As we have seen in Eq. (\ref{13prime}), $\theta_{13}^\prime$ should be about $1/5$ to generate the realistic $\theta_{13}$.
This can be achieved by further assuming $\Delta M \sim 1/5 \hspace{0.1cm} M_1$.
In summary, the following approximation can be taken to fit the experimental results,
\begin{equation}
6 y_4^2 \hspace{0.1cm} \sim \hspace{0.1cm}  3 y_5^{\prime 2} \hspace{0.1cm} \sim \hspace{0.1cm}  2 y_6^{\prime 2}, \hspace{1cm} M_1 \hspace{0.1cm}\sim \hspace{0.1cm} M_2 \hspace{0.1cm} \sim \hspace{0.1cm} 5 M_3 \hspace{0.1cm}\sim \hspace{0.1cm}5 |\Delta M|.
\label{approximation}
\end{equation}

\subsection{Leptogenesis}

Since the imaginary VEV of $\xi$ is the only source for CP violation in the lepton sector,
it is also expected to play an important role in the leptogenesis mechanism \cite{lepto},
which is a popular scenario for generating the baryon asymmetry of the universe
\begin{equation}\label{number}
\frac{n_{{\rm B}}}{s} = (8.79\pm 0.44) \times 10^{-11}.
\end{equation}
Here $n_{{\rm B}}$ and $s$ are the number densities of baryons and entropy respectively.
In this mechanism, the CP-violating and lepton-number-violating out-of-equilibrium decay of the lightest right-handed neutrino
creates the net lepton number $\Delta L$, which is then converted to baryon number by the sphaleron process \cite{sphaleron}.
To study this issue, right-handed neutrinos are transformed to the mass basis $N_{1}^\prime$,
$N_{2}^\prime$ and $N_{3}^\prime$ by a 1-3 rotation,
\begin{equation}
\left(
\begin{array}{ccc}
M_1^\prime&  &\\
 & M_2^\prime &\\
 & & M_3^\prime
\end{array}
\right)=
\left(
\begin{array}{ccc}
\cos{\theta}&  &\mp i\sin{\theta}\\
 & 1 &\\
\mp i\sin{\theta} & & \cos{\theta}
\end{array}
\right)
\left(
\begin{array}{ccc}
M_1&& \Delta M\\
&M_2&\\
\Delta M& &M_3
\end{array}
\right)
\left(
\begin{array}{ccc}
\cos{\theta}&  &\mp i\sin{\theta}\\
 & 1 &\\
\mp i\sin{\theta} & & \cos{\theta}
\end{array}
\right),
\end{equation}
with $\tan{2\theta}=2|\Delta M|/(M_1+M_3)$.
Correspondingly, the Yukawa couplings between $N_i^{\prime c}$ and $L_j$ are
\begin{equation}
\eta=
\left(
\begin{array}{ccc}
\cos{\theta}&  &\mp i\sin{\theta}\\
 & 1 &\\
\mp i\sin{\theta} & & \cos{\theta}
\end{array}
\right)\left(
\begin{array}{ccc}
2 y_4&\hspace{0.2cm} -y_4&\hspace{0.2cm}-y_4\\
y_5^\prime&\hspace{0.2cm}y_5^\prime & \hspace{0.2cm}y_5^\prime\\
0&\hspace{0.2cm}-y_6^\prime&\hspace{0.2cm} y_6^\prime
\end{array}
\right) \epsilon_2.
\end{equation}
And $\eta \eta^\dagger$ which will be needed below has a form as
\begin{equation}
\eta \eta^\dagger=\left(
\begin{array}{ccc}
6 y_4^2 \cos^2{\theta}+2 y_6^{\prime 2} \sin^2{\theta}& &i(\pm 6 y_4^2\mp 2 y_6^{\prime 2} )\cos{\theta}\sin{\theta}\\
 & 3y_5^{\prime 2}&\\
i(\mp 6 y_4^2\pm 2 y_6^{\prime 2} )\cos{\theta}\sin{\theta}& & 6 y_4^2 \sin^2{\theta}+2 y_6^{\prime 2} \cos^2{\theta}
\end{array}
\right) \epsilon_2^2.
\end{equation}

In the following analysis, we will adopt the approximation in Eq. (\ref{approximation}) which results in
\begin{equation}
M_1^\prime \sim M_2^\prime \sim 5 M_3^\prime, \hspace{1cm} \cos{\theta}\sim 1, \hspace{1cm} \sin{\theta}\sim 0.2.
\end{equation}
In this case, $N_3^{\prime}$ is the lightest right-handed neutrino and the CP asymmetry in its decay
to the lepton doublet $l_j$ can be expressed as \cite{flavoreffect}
\begin{equation}
\delta_{3}^{j} \hspace{0.1cm} \sim \hspace{0.1cm} \frac{-1}{8\pi}\hspace{0.1cm} \frac{1}{(\eta \eta^\dagger)_{33}}\hspace{0.1cm} {\rm{Im}}\hspace{0.1cm}\left\{\hspace{0.1cm}\eta_{1j}\hspace{0.1cm}\eta_{3j}^*\hspace{0.1cm}\left[\frac{3}{2}(\eta \eta^\dagger)_{13}\hspace{0.1cm}\frac{M_3^\prime}{M_1^\prime}+(\eta \eta^\dagger)_{31}\left(\frac{M_3^\prime}{M_1^\prime}\right)^2\hspace{0.1cm}\right]\hspace{0.1cm}\right\}\hspace{0.1cm}.
\end{equation}
The final baryon asymmetry can be estimated by
\begin{equation}\label{zero}
 \frac{n_{{\rm B}}}{s} = -c_{s}\sum_{j} \delta_{3}^{j}\hspace{0.1cm}\kappa.
\end{equation}
In this equation, $c_s \sim 10^{-3}$ is a coefficient accounting for the entropy dilution of $\Delta B$
and conversion factors in the sphaleron process.
$\kappa$ which has a value about $10^{-2}-10^{-3}$ is the efficiency factor due to the washout effects.
Although there is a CP phase in $\eta$, the total CP asymmetry is vanishing
\begin{equation}\label{sum}
  \sum_{j} \delta_{3}^{j}=\frac{-1}{8\pi}\hspace{0.1cm} \frac{1}{(\eta \eta^\dagger)_{33}}\hspace{0.1cm} {\rm{Im}}\hspace{0.1cm}\left\{\hspace{0.1cm}(\eta \eta^\dagger)_{13}\hspace{0.1cm}\left[\frac{3}{2}(\eta \eta^\dagger)_{13}\hspace{0.1cm}\frac{M_3^\prime}{M_1^\prime}+(\eta \eta^\dagger)_{31}\left(\frac{M_3^\prime}{M_1^\prime}\right)^2\hspace{0.1cm}\right]\hspace{0.1cm}\right\}\hspace{0.1cm}=0,
\end{equation}
leading to a zero baryon asymmetry.

However, if $N_3^\prime$ is lighter than $10^{12}$ GeV, the situation will change dramatically.
In this case, the Yukawa interaction of $\tau$ will approach equilibrium during the decay process,
making $\tau$ distinguishable from $\mu$ and $e$.
Thus, we have to take into consideration the flavor effects \cite{flavoreffect} which do not allow
the CP asymmetries for different flavors to be summed directly.
Alternatively, every $\delta_{3}^j$ should be with a corresponding efficiency factor $\kappa_{j}$
in the expression for baryon asymmetry
\begin{equation}
\frac{n_{{\rm B}}}{s} = -c_{s}\sum_{j} \delta_{3}^{j}\hspace{0.1cm}\kappa_{j},
\end{equation}
where $\kappa_{j}=\kappa/K_j$, with $K_j$ defined as
\begin{equation}\label{}
  K_j=\displaystyle \frac{\eta_{3j}\hspace{0.05cm}\eta_{3j}^*}{(\eta\eta^\dagger)_{33}}.
\end{equation}
From the experience of Eq. (\ref{zero}), we can see that the baryon asymmetry
will be proportional to $|\kappa_{\tau}-(1-\kappa_{\tau})|$, which can be approximated as $\kappa \sin^2{\theta}$.
At last, we can get the baryon asymmetry
\begin{equation}
\frac{n_{{\rm B}}}{s} \sim -c_{s}\hspace{0.1cm} \frac{-3}{16\pi}\hspace{0.1cm}  y^{ 2}\hspace{0.1cm}\epsilon_2^2 \hspace{0.1cm} \sin{\theta} \hspace{0.1cm}\frac{M_3^\prime}{M_1^\prime} \hspace{0.1cm} \kappa \hspace{0.1cm} \sin^{2}{\theta},
\end{equation}
where $y$ is used to denote all the Yukawa couplings and the $\mathcal O(1)$ coefficients have been omitted.
If we require $M_{3}^\prime$ to be in the range $10^{10}-10^{12}$ GeV and consider that
\begin{equation}\label{}
  m_3 \sim \frac{y^2 \hspace{0.1cm}\epsilon_2^2 \hspace{0.1cm} V_{\rm u}^2}{M_3^\prime},
\end{equation}
$y^2 \hspace{0.1cm}\epsilon_2^2$ should be about $10^{-4}-10^{-2}$.
As a consequence, the baryon asymmetry produced will be $10^{-14}-10^{-11}$,
consistent with the value given by Eq. (\ref{number}).

\subsection{Discussions}

Finally, several comments are given in order.

1. The first thing to note is the NLO corrections might disturb the mixing pattern, so we have to treat them carefully.
Due to the setting of this model, there are only two terms contributing to fermion masses at NLO,
\begin{equation}
y_7 N_1^c L H_{{\rm u}} \displaystyle \frac{\psi \omega}{\Lambda^2}+ y_8 N_3^c L H_{{\rm u}} \displaystyle \frac{\varphi \omega}{\Lambda^2}.
\end{equation}
After receiving this contribution, the Dirac neutrino mass matrix becomes
\begin{equation}
M_{{{\rm D}}}^\prime=\left(
\begin{array}{ccc}
2 y_4&\hspace{0.8cm} -y_4-\epsilon_1 y_7^\prime &\hspace{0.8cm} -y_4+ \epsilon_3 y_7^\prime  \\
  y_5^\prime &\hspace{0.8cm} y_5^\prime &\hspace{0.8cm}y_5^\prime \\
   2 \epsilon_3 y_8&\hspace{0.8cm} -y_6^\prime-\epsilon_3 y_8&\hspace{0.8cm} y_6^\prime- \epsilon_3 y_8
\end{array}
\right)\epsilon_2 V_{{\rm u}} ,
\end{equation}
where $\epsilon_3=V_5/\Lambda$ and $y_7^\prime=y_7\sqrt{(3\lambda_5 \lambda_{17}-3 \lambda_6 \lambda_{16})/(\lambda_7 \lambda_{17}-\lambda_6 \lambda_{18})}$.
The mass matrix for light neutrinos can be calculated in the usual way
\begin{equation}
M_{\nu}^{\prime \prime}= M_{{\rm D}}^{\prime {{\rm T}}} M_{{{\rm N}}}^{\prime -1} M_{{{\rm D}}}^{\prime}.
\end{equation}
As before, a TBM rotation is first performed,
\begin{equation}
\begin{array}{l}
\vspace{0.35cm}
U_{{\rm TBM}}^{{\rm T}} M_{\nu}^{\prime \prime} U_{{\rm TBM}}=
\left(
\begin{array}{ccc}
\sqrt{6} y_4& &\sqrt{6} \epsilon_3 y_8\\
 & \sqrt{3} y_5^\prime &\\
\sqrt{2}\epsilon_3 y_7^\prime & &\sqrt{2} y_6^\prime
\end{array}
\right)
M_{{\rm N}}^{\prime -1}
\left(
\begin{array}{ccc}
\sqrt{6} y_4&&\sqrt{2}\epsilon_3 y_7^\prime\\
 & \sqrt{3} y_5^\prime &\\
\sqrt{6} \epsilon_3 y_8 & &\sqrt{2} y_6^\prime
\end{array}
\right)(\epsilon_2 V_{{\rm u}} )^2=\\
\left(
\begin{array}{ccc}
\displaystyle\frac{6 (\epsilon_3^2 y_8^2M_1-2 \epsilon_3 y_4 y_8 \Delta M+y_4^2 M_3)}{D}& &\displaystyle\frac{ 2 \sqrt{3}(\epsilon_3 y_6^{\prime} y_8 M_1- y_4 y_6^{\prime} \Delta M+\epsilon_3 y_4 y_7^{\prime} M_3)}{D}\\
& \displaystyle\frac{3 y_5^{\prime 2}}{M_2}&\\
\cdots & &\displaystyle\frac{2 (y_6^{\prime 2} M_1-2 \epsilon_3 y_6^\prime y_7^\prime \Delta M+ \epsilon_3^2 y_7^{\prime 2}M_3)}{D}
\end{array}
\right)(\epsilon_2 V_{{\rm u}} )^2.
\end{array}
\end{equation}
This matrix is then diagonalized by $U(\theta_{13}^\prime)$ with
\begin{equation}
\begin{array}{llll}
\vspace{0.3cm}
\rho=0, & \hspace{0.3cm} \tan{2\theta_{13}^\prime}= \displaystyle\frac{2 \sqrt{3}( \epsilon_3 y_6^{\prime}y_8 M_1- y_4 y_6^\prime \Delta M+\epsilon_3 y_4 y_7^{\prime} M_3)}{y_6^{\prime 2}M_1+2 \epsilon_3 (3y_4 y_8-y_6^\prime y_7^\prime) \Delta M-3 y_4^2 M_3}, &\hspace{0.3cm}\rm{if} &\hspace{0.2cm}V_5=\pm \displaystyle \frac{M}{\sqrt{\lambda_4}};\\
\rho= \displaystyle\frac{\pi}{2}/\frac{3\pi}{2},  & \hspace{0.3cm} \tan{2\theta_{13}^\prime}=\displaystyle\frac{2 \sqrt{3}( |\epsilon_3| y_6^{\prime}y_8 M_1- y_4 y_6^\prime |\Delta M|+|\epsilon_3| y_4 y_7^{\prime} M_3)}{y_6^{\prime 2}M_1-2 \epsilon_3 (3y_4 y_8+y_6^\prime y_7^\prime ) \Delta M+3 y_4^2 M_3}, & \hspace{0.3cm}\rm{if}&\hspace{0.2cm}V_5=\pm i \displaystyle \frac{M}{\sqrt{\lambda_4}}.
\end{array}
\end{equation}
Thus, we can say the mixing pattern is stable against at least NLO corrections.

2. If the requirement of CP symmetry is relaxed, this model can give the TM2 mixing with an arbitrary $\rho$,
which is determined by the diagonalization of Eq. (\ref{theta13})
in which $y_4$, $y_5^\prime$ and $y_6^\prime$ are complex coefficients for now.
In this case, $\theta_{23}$ can be obtained through Eq. (\ref{theta23}).
For example, if we take $\rho$ (which approximates to $\delta$ in the standard parameterization)
as the best-fit value $1.34 \pi$ in TABLE \Rmnum{1}, $\sin^2{\theta_{23}}$ will be 0.55 and close to the best-fit value 0.567.

3. Analogously, the model building and phenomenological analysis in this section completely apply to the TM1 mixing \cite{tm1}
\begin{equation}
U_{{\rm PMNS}}\hspace{0.1cm}=
\left(
\begin{array}{ccc}
\vspace{0.2cm}
\displaystyle\frac{\sqrt 2}{\sqrt 3}&\displaystyle\frac{1}{\sqrt 3}&0\\
\vspace{0.2cm}
-\displaystyle\frac{1}{\sqrt 6}&\displaystyle\frac{1}{\sqrt 3}&\displaystyle\frac{1}{\sqrt 2}\\
\vspace{0.2cm}
\displaystyle\frac{1}{\sqrt 6}&-\displaystyle\frac{1}{\sqrt 3}&\displaystyle\frac{1}{\sqrt 2}
\end{array}
\right)
\left(
\begin{array}{ccc}
\vspace{0.2cm}
1&  &\\
\vspace{0.2cm}
&\cos{\theta_{23}^\prime}&\sin{\theta_{23}^\prime}e^{-i \rho}\\
\vspace{0.2cm}
& -\sin{\theta_{23}^\prime}e^{i \rho}&\cos{\theta_{23}^\prime}
\end{array}
\right).
\label{tm1}
\end{equation}
Put simply, if we assign $\xi$ the quantum numbers of $Z_2^2$ and $Z_2^3$,
its VEV will introduce the mixing between $N_2$ and $N_3$.
That is to say, the PMNS matrix would be the TBM matrix multiplied by a 2-3 rotation from the right-hand side, just like the TM1 mixing.

4. The last point to stress is that the CP symmetry imposed on the model is consistent with the flavor symmetry $A4$,
although it is not defined in the way as the so-called generalized CP transformation does.
This is because in the particular basis of $A4$ we have chosen \cite{a4}, the would-be generalized CP transformation
can be represented as the identity matrix times a phase, i.e., a trivial one \cite{lindner}.
In other words, the naive CP transformation $\phi\rightarrow \phi^*$ can work well in our model.

\section{Summary}

In this paper, we have attempted to modify the TBM mixing as minimally as possible
to accommodate the recent observation of a relatively large $\theta_{13}$.
Above all, we have examined whether there is a neutrino mass matrix with a simple form
that can describe the realistic mixing scheme.
By analogy with that for the TBM mixing, we find one neutrino mass matrix with only three independent parameters,
which is connected with the Friedberg-Lee symmetry and $\mu-\tau$ symmetry breaking.
Unlike in the TBM case, the masses and mixing angles of this mass matrix are correlated,
so that the mass values can be determined with reference to the values of mixing angles.
The values of mass sum and $m_{\beta\beta}$ are close to the experimental sensitivities,
thus will be observed or excluded in a short time.

As a matter of fact, the mixing pattern given by this mass matrix is the so-called TM2 mixing.
Therefore, we have also discussed the way to build a model generating this mixing pattern,
invoking the minimal modification to a model that produces the TBM mixing \cite{example}.
The model is constructed with $A_4 \times U(1)\times Z_2^1 \times Z_2^2 \times Z_2^3$ as the flavor symmetries.
The mass matrix for charged leptons is diagonal and the hierarchies among them are guaranteed by the $U(1)$ symmetry.
Similarly, the mass matrix for right-handed neutrinos is also diagonal, because of the $Z_2^1 \times Z_2^2 \times Z_2^3$ symmetry.
Their Yukawa couplings with left-handed neutrinos have a special form,
as a consequence of the specific VEV alignments, which are justified to NLO, possessed by the flavon fields.
At this stage, the mixing pattern for light neutrinos is the TBM after the seesaw mechanism.
However, a flavon field which acquires a VEV can introduce the mixing between the first and third right-handed neutrinos,
transforming the mixing pattern to the TM2.
More interestingly, this VEV can be purely imaginary if we impose the CP symmetry on this model, 
leading to the maximal CP violation and maximal $\theta_{23}$.
On the other side, this maximal CP violation gives a zero total CP asymmetry in the leptogenesis mechanism.
But after the flavor effects are considered, the observed value of baryon asymmetry can be marginally reproduced.
If the CP symmetry is given up, the CP phase will become free and $\theta_{23}$ can departure from $45^\circ$.
Besides, it is found the NLO contributions do not change the mixing pattern.
The last thing to mention is that this approach can be directly generalized to realizing the TM1 mixing.

In conclusion, the TBM mixing can be modified in a minimal way, in terms of both the mass matrix and model building,
to accommodate the non-zero $\theta_{13}$ and CP phase.
Therefore, it can still serve as a guide for model building.

\begin{acknowledgments}
I am very grateful to Professor Zhi-zhong Xing for useful comments and suggestions and Professor Chun Liu for his support.
I am also indebted to Jue Zhang and Shun Zhou for their help with the flavor effects in leptogenesis and Ye-ling Zhou for helpful discussions
about generalized CP transformation.
This work was supported in part by the National Natural Science
Foundation of China under Nos. 11375248 and 11135009.

\end{acknowledgments}

\end{document}